\documentclass[conference]{IEEEtran}

\makeatletter
\newcommand\semihuge{\@setfontsize\semihuge{22.3}{22}}
\makeatother

\usepackage{algpseudocode}
\usepackage{algorithm}
\usepackage{algorithmicx}

\usepackage{lipsum} 
\usepackage{arydshln}
\usepackage[dvips]{color}
\usepackage{comment}
\usepackage{todonotes}
\usepackage{epsf}
\usepackage{epsfig}
\usepackage{times}
\usepackage{epsfig}
\usepackage{graphicx}
\usepackage{bbold}
\usepackage{mathtools}
\usepackage{mathrsfs}
\usepackage{amssymb}
\usepackage{pdfpages}
\usepackage{epstopdf}
\newfloat{algorithm}{t}{lop}

\usepackage{amsmath}
\usepackage{dsfont}
\usepackage{lettrine} 
\usepackage{amsmath,epsfig,amssymb,algorithm,algpseudocode,amsthm,cite,url}
\usepackage{subcaption}
\allowdisplaybreaks
\usepackage{csquotes}

\usepackage{verbatim}
\usepackage[english]{babel}
\usepackage{hyperref}
\usepackage{amsmath,amssymb}

\usepackage[justification=centering]{caption}
\usepackage{verbatim}

\newtheorem{definition}{\bf Definition}

\begin{document}
	%
	\title{Game Theory for Secure Critical Interdependent Gas-Power-Water Infrastructure\vspace{-0.4cm}}
	%
	%
	\IEEEoverridecommandlockouts
\author{\IEEEauthorblockN{Aidin Ferdowsi\IEEEauthorrefmark{1}, Anibal Sanjab\IEEEauthorrefmark{1}, Walid Saad\IEEEauthorrefmark{1}, Narayan B. Mandayam\IEEEauthorrefmark{2}}
	\IEEEauthorblockA{\IEEEauthorrefmark{1}
		Wireless@VT, Bradley Department of Electrical and Computer Engineering, Virginia Tech, Blacksburg, VA, USA,\\
		Emails: \{aidin,anibals,walids\}@vt.edu}
		\IEEEauthorblockA{\IEEEauthorrefmark{2}
		WINLAB, Dept. of ECE, Rutgers University, New Brunswick, NJ, USA,		Email:  narayan@winlab.rutgers.edu}\vspace{-1cm}
		\thanks{This research was supported by the U.S. National Science Foundation under Grants ACI-1541105, ACI-1541069, and CNS-1446621.\vspace{-0.5cm}}
}%

	\maketitle
	
\begin{abstract}
A city's critical infrastructure such as gas, water, and power systems, are largely interdependent since they share energy, computing, and communication resources. This, in turn, makes it challenging to endow them with fool-proof security solutions. In this paper, a unified model for interdependent gas-power-water infrastructure is presented and the security of this model is studied using a novel game-theoretic framework. In particular, a zero-sum noncooperative game is formulated  between a malicious attacker who seeks to simultaneously alter the states of the gas-power-water critical infrastructure to increase the power generation cost and a defender who allocates communication resources over its attack detection filters in local areas to monitor the infrastructure. At the mixed strategy Nash equilibrium of this game, numerical results show that the expected power generation cost deviation is 35\% lower than the one resulting from an equal allocation of resources over the local filters. The results also show that, at equilibrium, the interdependence of the power system on the natural gas and water systems can motivate the attacker to target the states of the water and natural gas systems to change the operational states of the power grid. Conversely, the defender allocates a portion of its resources to the water and natural gas states of the interdependent system to protect the grid from state deviations.
\end{abstract}\vspace*{-0.2cm}

	%
	\IEEEpeerreviewmaketitle

\vspace*{-0.1 cm}	
\section{Introduction}\vspace*{-0.12 cm}
Owing to the integration of information and communication technologies, critical infrastructure such as electric power grids, water systems, and natural gas distribution systems will be largely interdependent \cite{fagiani2014computational,farhangi2010path,rinaldi2001identifying}. This critical infrastructure interdependence is identified in\cite{rinaldi2001identifying} in which the authors describe the interdependence between the functionality of various systems such as the water, natural gas, transportation, and telecommunication networks. Considering these infrastructure as cyber-physical systems (CPSs) where the physical network is monitored by an intelligent control layer \cite{shi2011survey,mo2012cyber}, their interdependence makes the design of accurate monitors of the system states challenging.
	
The security of critical cyber-physical infrastructure in the face of cyber and physical attacks has received increasing attention, recently. In \cite{pasqualetti2013attack,fawzi2014secure,liu2011false}, several control-theoretic solutions have been introduced for securing CPSs. In \cite{pasqualetti2013attack}, the authors analyzed the impact of several additive attacks on a linear state space CPS model. Secure control design for CPSs is addressed in \cite{fawzi2014secure} which presents a resilient state estimation technique in the presence of attacks on the actuators of the system. The work in \cite{liu2011false} analyzed the impact of cyber attacks against the
state estimators where it was shown that attacks can be initiated even
in presence of strict limitations on the attacker’s resources. 

In addition, a number of recent works studied the game-theoretic security of critical infrastructure such as in \cite{zhu2011stackelberg,gupta2014three,ferdowsi2017colonel}. A Stackelberg game was proposed in \cite{zhu2011stackelberg} to maintain the performance of a control system despite the presence of attack. In \cite{gupta2014three} and \cite{ferdowsi2017colonel}, the problem of security resource allocation for CPS is studied using a Colonel Blotto game framework that yields an optimal allocation of limited defense resources over the various CPS nodes.  However, \cite{zhu2011stackelberg,gupta2014three,ferdowsi2017colonel} do not account for the interdependence between critical infrastructure in optimal allocation of defense resources, while \cite{pasqualetti2013attack,fawzi2014secure,liu2011false} do not consider the limitations on the resources of the defender when monitoring the system. \vspace*{-0.1 cm}

The main contribution of this paper is a novel framework for analyzing the interdependence between gas, power, and water critical infrastructure in presence of a malicious adversary. In particular, we first present a unified model for the interdependent gas-power-water infrastructure. Then we study the optimal allocation of defense resources over the subsystems of interdependent power, water, and natural gas infrastructure systems by taking into account their synergies. We formulate the problem as a two-player noncooperative zero-sum game between the owner of the interdependent gas-power-water infrastructure, acting as defender, and an adversary. In this game, the defender needs to allocate communication resources across the different subsystems of the interconnected infrastructure for monitoring purposes which in turn lead to improved detection and enabling of protective actions. Meanwhile, the attacker seeks to disrupt and deviate the entire system from its operational state by simultaneously choosing a set of entities over which it will initiate state attacks. One key property of this game is that it allows identification of the behavior of the defender and attacker in presence of interdependent infrastructure. We show how the interconnectivity between critical infrastructure can exacerbate security risks, even though it achieves better performance and operational quality. Numerical results show how an adversary can benefit from this interdependence to induce damage on the power system by launching an attack on components of the water or gas systems. The results also show that a game-theoretic approach for communication resource allocation yields better control on the system operation under attack, compared to an equal allocation of resources.\vspace*{-0.1 cm}

The rest of the paper is organized as follows. Section \ref{sysmod} presents the state space model for the interdependent critical infrastructure. In Section \ref{attackmod}, the attack model is presented. Section \ref{defmod} presents system monitors which are guaranteed to detect state attacks. In Section \ref{gamemod}, the proposed game-theoretic framework is presented. Simulation results are analyzed in Section \ref{sim results} while conclusions are drawn in Section \ref{conclusion}.
\section{System Model}\label{sysmod}
Consider a smart city that encompasses various interconnected CPSs, namely, electric, natural gas, and water distribution systems whose interdependence and synergy enable a smart and sustainable operation of the entire city. To formally define such infrastructure, next, we introduce a mathematical state space representation which captures their interdependence. This proposed analytical framework enables an accurate understanding of the dynamic operation of the considered critical infrastructure which allows analysis of their security. \vspace{-0.185 cm}
\subsection{Electric Power Systems}
The electric power system is composed of generators and loads interconnected via transmission lines and substations. The dynamics of a generator are dependent on the input mechanical power received from burning fuel (such as natural gas or coal) or flowing wind or water, and on the output power which it must supply to meet the city's demand. Hence, considering static frequency-dependent loads at the time-scale of our analysis, we can model the power system dynamics following the linear swing model \cite{scholtz2004observer}:
\begin{align}\centering \label{elecDAE}
\begin{bmatrix}
\boldsymbol{I} & 0 & 0 \\
0 & \boldsymbol{M} & 0 \\
0 & 0 & 0
\end{bmatrix} \hspace{-0.2cm}
\begin{bmatrix}
\dot{\boldsymbol{\delta}}  \\
\dot{\boldsymbol{\omega}} \\
\dot{\boldsymbol{\theta}} 
\end{bmatrix}\hspace{-0.15cm}
=\hspace{-0.05cm} - \begin{bmatrix}
0 		& -\boldsymbol{I}  & 0 
\\
\boldsymbol{L}_{gg}  & \boldsymbol{D} &\boldsymbol{L}_{gl}\\
\boldsymbol{L}_{lg}  & 0   &\boldsymbol{L}_{ll}
\end{bmatrix}\hspace{-0.2cm}
\begin{bmatrix}
\boldsymbol{\delta}  \\
\boldsymbol{\omega} \\
\boldsymbol{\theta} 
\end{bmatrix}\hspace{-0.1cm}
+\hspace{-0.1cm}
\begin{bmatrix}
0  \\
\boldsymbol{P}_g \\
\boldsymbol{P}_l 
\end{bmatrix},
\end{align}
where $ \boldsymbol{\delta} $ and $ \boldsymbol{\omega} $ are, respectively, the vectors of rotors' phase angles (i.e. angular displacement) and angular speeds which constitute the state variables of the electric system. $ \boldsymbol{L}_{gg} $, $ \boldsymbol{L}_{gl} $, $ \boldsymbol{L}_{lg} $, and $ \boldsymbol{L}_{ll} $ are the susceptance matrices of the system where subscripts $ l $ and $ g $ stand for generator and load, respectively, $ \boldsymbol{M} $ and $ \boldsymbol{D} $ are diagonal matrices representing, respectively, the generators' inertia constants and damping coefficients, while $ \boldsymbol{P}_g $ and $ \boldsymbol{P}_l $ are the vectors of net power injected at the generator and load buses, respectively. \vspace{-0.185 cm}
\subsection{Natural Gas and Water Systems} 
The natural gas distribution system is composed of various interconnected components including supply units, storage units, pipelines, compressors, and loads. Natural gas is supplied from gas wells, commonly located at remote sites \cite{liu2009security} and which can be modeled as positive injections to the natural gas system with constant head pressures. In addition to gas wells, natural gas storage units are distributed in the network to provide gas to the system during times of high demand and receive gas from the system when demand is low. Thus, the head pressure of the storage units can be modeled as:
\begin{align}
R_r\dot{h}_r=\sum_{j \in \mathcal{N}^\textrm{in}_r}Q_{js}-\sum_{k \in \mathcal{N}^\textrm{out}_s}Q_{ri},
\end{align}
where $ h_r $ is the head pressure of the storage junction, $ Q_{jr} $ is the flow between junctions $ j $ and junction $ r $, $ \mathcal{N}_r^\textrm{in} $ and $ \mathcal{N}_r^\textrm{out} $ are the set of junctions which have positive or negative flow to junction $ r $, respectively, and $ R_r $ is the charging ratio of the corresponding gas tank.

The flow in a pipeline depends on the pressure drop along the pipe\cite{liu2009security} and is modeled as:\vspace{0.15cm}
\begin{align}\label{pipeflow}
Q_{ij}&=\mathrm{sgn}(h_i,h_j)C_{ij}\sqrt{|h_i^2-h_j^2|},
\end{align}
where $ \mathrm{sgn}(h_i,h_j) =1$ if $ h_i \geq h_j $ and  $ \mathrm{sgn}(h_i,h_j) =-1$ if $ h_i < h_j $, and $ C_{ij} $ is a pipeline constant which depends on the physical characteristics of the pipeline, the environment, and the gas composition. The pressure loss in pipelines can be compensated by compressors. In this respect, the flow in the gas pipeline can then be represented as follows:\vspace{0.15cm}
\begin{align}\label{compressorflow}
Q_{ij}=\mathrm{sgn}(h_i,h_j) \frac{P_c}{k_{2}-k_{1}\left[\frac{\max(h_i,h_j)}{\min(h_i,h_j)}\right]^{\alpha}},
\end{align}
where $ P_c $ is the power demand of the compressor, while $ k_{1} $, $ k_{2} $, and $ \alpha $ are design parameters of the compressor\cite{liu2009security}.

\emph{Loads:} In natural gas systems, the demand junctions are typically modeled as follows \cite{liu2009security}:\vspace*{0.15cm}
\begin{align}
d_i= \sum_{j \in \mathcal{N}^\textrm{in}_i}Q_{ji}-\sum_{k \in \mathcal{N}^\textrm{out}_i}Q_{ik},
\end{align}
where $ d_i $ is the static demand at junction $ i $.

By using an analogy to the gas system, we can model flow equation in the water network as follows \cite{pasqualetti2013attack}:
\begin{align}\label{waterpipeflow}
Q_{ij}=\mathrm{sgn}(h_i,h_j)C_ij|h_i-h_j|^{\frac{1}{1.85}}.
\end{align}  
Also, treatment plants for purification and pressure control inside the water system require electric power and they make the water system dependent on the electric system \cite{Stillwell2010water}.\vspace{-0.23 cm}
\subsection{Interconnected Critical Infrastructure Model}
Various components in the gas, power, and water systems can be largely interdependent. For instance, many generators can be supplied by natural gas and most generators can use water at different steps of power generation such as from steam condensation and temperature control \cite{Macknick2012water}. These large water and gas requirements imply that the generators will constitute demand junctions in the water and gas systems. In our model, we categorize the generators into two groups: a) a first group in which generators are supplied by natural gas and the mechanical input power, $ P_l $, of these generators is proportional to the input natural gas, $ d_i $, and b) a second group of generators that are nuclear, coal supplied, or wind turbines. Additionally, the power generated at all generators is proportional to the input water. Furthermore, the water treatment plants and gas compressors constitute one of the major loads of the power system. Therefore, the interdependence of the power system on water and natural gas systems can be captured as follows:\vspace*{0.3 cm}
\begin{align}\label{powertogaswater}
	\left[\begin{array}{c}
	\boldsymbol{P}_g \\
	\boldsymbol{P}_l \\
	\end{array}\right]=	\left[\begin{array}{c c}
	\boldsymbol{C}'^G & \boldsymbol{C}'^W 
	\end{array}\right]
	\left[\begin{array}{c}
	\boldsymbol{h}_S^G \\
	\boldsymbol{h}_J^G \\\hdashline
	\boldsymbol{h}_S^W \\
	\boldsymbol{h}_J^W
	\end{array} \right],
\end{align}
\begin{figure*}[!t]
	\begin{equation}\label{GeneralDAE}
	\left[
	\begin{array}{c : c : c}
	\begin{matrix}
	\boldsymbol{I} & \boldsymbol{O} & \boldsymbol{O} \\
	\boldsymbol{O} & \boldsymbol{M} & \boldsymbol{O} \\
	\boldsymbol{O} & \boldsymbol{O} & \boldsymbol{O}
	\end{matrix} & \boldsymbol{O}  & \boldsymbol{O} \\
	\hdashline
	\boldsymbol{O} & 	\begin{matrix}\boldsymbol{R}_c & \boldsymbol{O}\\
	\boldsymbol{O} &\boldsymbol{O}			
	\end{matrix} & \boldsymbol{O} \\
	\hdashline
	\boldsymbol{O} & \boldsymbol{O} &		\begin{matrix}\boldsymbol{R}_t & \boldsymbol{O}\\
	\boldsymbol{O} &\boldsymbol{O}			
	\end{matrix}
	\end{array}
	\right]
	\left[
	\begin{array}{c}
	\dot{\boldsymbol{\delta}}  \\
	\dot{\boldsymbol{\omega}}_G \\
	\dot{\boldsymbol{\omega}}\\
	\dot{\boldsymbol{\theta}}_c\\
	\dot{\boldsymbol{\theta}}_{t}\\
	\dot{\boldsymbol{\theta}}\\\hdashline
	\dot{\boldsymbol{h}}_s^G\\
	\dot{\boldsymbol{h}}_j^G\\\hdashline
	\dot{\boldsymbol{h}}_s^W\\
	\dot{\boldsymbol{h}}_j^W
	\end{array}
	\right]
	=
	\hspace*{-0.05cm}
	\left[
	\begin{array}{c c c:c:c}
	\boldsymbol{O} & \boldsymbol{I} & \boldsymbol{O} & \boldsymbol{O} & \boldsymbol{O}\\
	-\boldsymbol{L}_{gg} & \boldsymbol{D} & \boldsymbol{L}_{gl} &\boldsymbol{C}^G & \boldsymbol{C}^W \\
	-\boldsymbol{L}_{lg} & \boldsymbol{O} & \boldsymbol{L}_{ll}& \boldsymbol{O} & \boldsymbol{O}\\\hdashline
	& \boldsymbol{O} & &\boldsymbol{C}^G & \boldsymbol{O}\\\hdashline
	& \boldsymbol{O} & & \boldsymbol{O} & \boldsymbol{C}^W
	\end{array}
	\right]
	\hspace*{-0.1cm}
	\begin{bmatrix}
	\boldsymbol{\delta}  \\
	\boldsymbol{\omega}_G \\
	\boldsymbol{\omega}\\
	\boldsymbol{\theta}_c\\
	\boldsymbol{\theta}_{t}\\
	\boldsymbol{\theta}\\\hdashline
	\boldsymbol{h}_s^G\\
	\boldsymbol{h}_j^G\\\hdashline
	\boldsymbol{h}_s^W\\
	\boldsymbol{h}_j^W
	\end{bmatrix},
	\vspace*{-0.7cm}
	\end{equation}
\end{figure*}where $ \boldsymbol{h}_s^G $ is the vector of the head pressure of the gas storage units having $ n_s^G $ elements, $ \boldsymbol{h}_s^W $ is the vector of the head pressure of the water storage units having $ n_s^W $ elements, $ \boldsymbol{h}_j^G $ is the vector of the head pressure at the gas junctions having $ n_j^G $ elements, and $ \boldsymbol{h}_j^W $ is the vector of head pressure at the water junctions having $ n_j^W $ elements. The characteristics of pipelines, treatment centers, and  compressors are captured in the $ (n_g^G+n_g+n_c+n_t) \times  (n_s^G+n_j^G)$ matrix $ \boldsymbol{C}^{'G} $ and $ (n_g^G+n_g+n_c+n_t) \times (n_s^W+n_j^W) $ matrix $ \boldsymbol{C}^{'W}$, respectively. 

By modifying the rows and columns of $ \boldsymbol{C}^{'G} $ and $ \boldsymbol{C}^{'W} $, we can substitute \eqref{powertogaswater} into \eqref{elecDAE} and derive a general differential algebraic state space model for the interconnected infrastructure as in \eqref{GeneralDAE}. In \eqref{GeneralDAE}, $ \boldsymbol{\delta} $ is the $ n_g^G+n_g \times 1 $ vector of generator phase angles, $ \boldsymbol{\omega}_G $ is the $ n_g^G \times 1 $ vector of angular speeds of generators running on natural gas, $ \boldsymbol{\omega} $ is the $ n_g \times 1 $ vector of angular speeds of generators running on resources other than natural gas, $ \boldsymbol{\theta}_c $ is the $ n_c \times 1$ vector of voltage phase angles at buses supplying power to gas compressors, $ \boldsymbol{\theta}_t $ is the $n_t \times 1$ vector of voltage phase angles at buses supplying power to water treatment plants, $ \boldsymbol{\theta} $ is the $ n_l \times 1$ vector of voltage phase angles at other types of buses, $ \boldsymbol{h}_s^G $ is the $ n_s^G \times 1$ vector of the head pressure of the gas storage units, $ \boldsymbol{h}_s^W $ is the $ n_s^W \times 1 $ vector of the head pressure of the water storage units, $ \boldsymbol{h}_j^G $ is the $ n_j^G \times 1 $ vector of head pressure at the junctions in the gas network, and $ \boldsymbol{h}_j^W $ is the vector of the $ n_j^W \times 1$ head pressure at the junctions of the water system. Here, all the vectors capture the states of the interconnected system. The dependence of the electric system on natural gas and water systems is captured by matrices $ \boldsymbol{C}^G $ and $ \boldsymbol{C}^W$, respectively. $ \boldsymbol{I} $ and $ \boldsymbol{O} $ are, respectively, the identity matrix and zero matrix with proper dimensions. Equation \eqref{GeneralDAE} can be rewritten as follows:
\begin{equation}\label{interDAE}\vspace*{-0.2cm}
\boldsymbol{E}\dot{\boldsymbol{x}}=\boldsymbol{A}\boldsymbol{x},
\end{equation} 
where $ \boldsymbol{x}\in \mathbb{R}^n $ is the state vector, $ \boldsymbol{A} \in \mathbb{R}^{n \times n} $, and the diagonal matrix $ \boldsymbol{E} \in \mathbb{R}^{n \times n}  $ is a constant matrix that reflects the system characteristics.
In addition, in CPSs, a set of sensors are spread around the system to collect measurements and report them to the administrator of the interdependent system.  We assume the owners of all three infrastructure as a single administrator since they work together to control their infrastructure. The sensor outputs are related to the system states following the output equation:\vspace{-0.2cm}
\begin{align}
	\boldsymbol{y}&=\boldsymbol{C}\boldsymbol{x},
\end{align}
where $ \boldsymbol{y} \in \mathbb{R}^p $ represents the sensor data vector and $\boldsymbol{C} \in \mathbb{R}^{p \times n}$ is known as the output matrix.
This large-scale interdependence among the dynamic states of the electric, water, and natural gas systems makes the resulting critical infrastructure vulnerable to state attacks as explained next. 
\section{Attack Model}\label{attackmod}\vspace*{-0.15cm}
State attacks on the considered critical interdependent gas-power-water infrastructure can be modeled as an additive attack to the descriptor system in \eqref{GeneralDAE}. Such attacks will lead to deviations in the states and sensor outputs. The dynamic system under state attack will then be given by\cite{pasqualetti2013attack}:\vspace*{-0.05cm}
\begin{equation}
	\begin{aligned}
		\label{attackdiff}
		\boldsymbol{E}\dot{\tilde{\boldsymbol{x}}}&=\boldsymbol{A}\tilde{\boldsymbol{x}}+\boldsymbol{b}v,\\
		\tilde{\boldsymbol{y}}&=\boldsymbol{C}\tilde{\boldsymbol{x}},
	\end{aligned}
\end{equation}
where $ \tilde{\boldsymbol{x}}\in \mathbb{R}^n $ and $ \tilde{\boldsymbol{y}} \in \mathbb{R}^p $ are the state variables of the system state and measurements in the presence of a state attack. $ v\in \mathbb{R} $ is the state attack value, $ \boldsymbol{b} \in \mathbb{R}^n$ is a vector containing $n$ elements only $K$ of which are nonzero having a value equal to 1. These nonzero elements correspond to the states selected by the attacker to initiate a state attack. We define $ \kappa $ to be the set of indices of the nonzero elements of vector $ \boldsymbol{b} $.	
In this regard, to capture the effect of the attack on the state variables, we apply the Laplace transform to the difference between  \eqref{interDAE} and \eqref{attackdiff}, which yields:\vspace*{-0.15cm}
\begin{align}\label{n-tuple}
\left(s\boldsymbol{E}-\boldsymbol{A}\right)\Delta \boldsymbol{x}(s)=\boldsymbol{b}v(s),
\end{align}
where $ \Delta \boldsymbol{x}(s) $ is the vector of state deviation, in the Laplace domain, in the presence of an attack. \eqref{n-tuple} is a system of $ n $ equations, therefore, assuming that our system is originally stable (under no attack), $ s\boldsymbol{E}-\boldsymbol{A}  $ can be considered to be invertible and, then, the solution of system \eqref{n-tuple} can be obtained using Cramer's rule as follows:\vspace*{-0.15cm}
\begin{align}\label{solution}
\Delta x_i(s)=\frac{|(s\boldsymbol{E}-\boldsymbol{A})_{i,v(s)}|}
{|s\boldsymbol{E}-\boldsymbol{A}|},
\end{align}
where $ |s\boldsymbol{E}-\boldsymbol{A}| $ is the determinant of matrix $ s\boldsymbol{E}-\boldsymbol{A} $ and $|(s\boldsymbol{E}-\boldsymbol{A})_{i,v(s)}|$ is the determinant of matrix $s\boldsymbol{E}-\boldsymbol{A}$ when its $i$-th column is replaced by $ \boldsymbol{b}v(s) $. The numerator of the fraction in \eqref{solution} can be written as:\vspace*{-0.1cm}
\begin{align}
|(s\boldsymbol{E}-\boldsymbol{A})_{i,v(s)}|=\sum_{j \in \kappa} (-1)^{i+j}|(s\boldsymbol{E}-\boldsymbol{A})_{\bar{i},\bar{j}}|v_j(s),
\end{align}
where $|(s\boldsymbol{E}-\boldsymbol{A})_{\bar{i},\bar{j}}|$ is the determinant of matrix $ s\boldsymbol{E}-\boldsymbol{A} $ excluding its $ i $-th column and $ j $-th row. Hence, we can simplify \eqref{solution} as follows:
\begin{align}\vspace{-0.2cm}
\Delta x_i(s)=\sum_{j \in \kappa} 	(-1)^{i+j}\frac{|(s\boldsymbol{E}-\boldsymbol{A})_{\bar{i},\bar{j}}|}{|s\boldsymbol{E}-\boldsymbol{A}|}v_j(s).
\end{align}
As such, we can quantify \emph{the deviation that is caused by the attacker on any system state} $ i $ in the time domain as follows:\vspace*{-0.05cm}
\begin{align}\label{attackinfluence}
\Delta x_i(t)=\sum_{j \in \kappa}(-1)^{i+j}\mathscr{L}^{-1}\left\{ 	\frac{|(s\boldsymbol{E}-\boldsymbol{A})_{\bar{i},\bar{j}}|}{|s\boldsymbol{E}-\boldsymbol{A}|}v_j(s)\right\},
\end{align}where $ \mathscr{L}^{-1}\{\} $ is the Laplacian inverse transform.
Here, we assume that the cost of power generation can be expressed as a function of the system states \cite{Weron2014} -- linearized around a certain operating state, using a first order Taylor series approximation -- and, hence, the cost deviation will be given by:
 \begin{align}
 	\Delta p=\sum_{i \in \mathcal{N}_e} c_{p_i}|\Delta x_i|,
 \end{align}
where $ \Delta p $ is the deviation in the real-time power generation cost, $ \mathcal{N}_e $ is the set of states inside the power system, and $ c_{p_i} $ is the portion of the effect of the deviation of state $ i $ on the cost, which is derived from linearization. Finally, the variation of power generation cost that is caused by the attack vector $ \boldsymbol{b}v $ for a duration of $ t_a $ seconds can be written as follows:
\begin{equation}
	\begin{aligned}\label{pricechange}
		\Delta p(t_a&,\kappa)=\int_{0}^{t_a}\sum_{i\in \mathcal{N}_e}c_{p_i}\times\\
		&\left|\sum_{j \in \kappa}(-1)^{i+j}\mathscr{L}^{-1}\left\{ 	\frac{|(s\boldsymbol{E}-\boldsymbol{A})_{\bar{i},\bar{j}}|}{|s\boldsymbol{E}-\boldsymbol{A}|}v_j(s)\right\}\right|dt.
	\end{aligned}
\end{equation}
We assume that the attacker can benefit from the deviation in the real-time cost of electric power generation and, therefore, seeks to maximize \eqref{pricechange} by initiating attacks on a properly selected set of states. In response, the defender can decrease the deviation in the cost by reducing the available time for the attacker to stay undetectable in the system. \vspace*{-0.3 cm}
\section{Attack Detection} \label{defmod} \vspace*{-0.1 cm}
Consider the descriptor system \eqref{attackdiff} for a known initial system state $ \boldsymbol{x}(0) $. In this case, an attack detection filter that is guaranteed to detect the presence of attacks \cite{pasqualetti2013attack} will be:
\begin{equation}
	\begin{aligned}
		\boldsymbol{E}\dot{\boldsymbol{z}}&=(\boldsymbol{A}+\boldsymbol{G}\boldsymbol{C})\boldsymbol{z}-\boldsymbol{G}\boldsymbol{y},\\
		\boldsymbol{r}&=\boldsymbol{C}\boldsymbol{z}-\boldsymbol{y},
	\end{aligned}
\end{equation}
where $ \boldsymbol{z}(0)=\boldsymbol{x}(0) $ and block diagonal matrix $ \boldsymbol{G} \in \mathbb{R}^{n \times p} $ is such that the pair $ (\boldsymbol{E},\boldsymbol{A}+\boldsymbol{G}\boldsymbol{C}) $ is regular and Hurwitz. Then $ \boldsymbol{r}(t)=0 $ at all times $ t \in \mathbb{R}_{\geq 0} $ if and only if $ \boldsymbol{b}v = 0 $ at  all times $ t \in \mathbb{R}_{\geq 0} $ and, in the absence of attacks, the filter error $ \boldsymbol{z}-\boldsymbol{x} $ is exponentially stable\cite{pasqualetti2013attack}. To implement this approach, we need $ N $ \emph{disjoint subsystems of the system} \eqref{attackdiff} with $ n_i $ state variables in each subsystem $ i $. Here, $ \boldsymbol{E} $ and $ \boldsymbol{C} $ are block-diagonal. To capture these subsystems, we can write the matrix $ \boldsymbol{A} $:
\begin{align}
\boldsymbol{A}=
\begin{bmatrix}
\boldsymbol{A}_1 & \cdots & \boldsymbol{A}_{1N}\\
\vdots & \vdots & \vdots \\
\boldsymbol{A}_{N1}& \cdots & \boldsymbol{A}_N
\end{bmatrix}=\boldsymbol{A}_D+\boldsymbol{A}_C,
\end{align}
where $ \boldsymbol{A}_i \in \mathbb{R}^{n_i \times n_i} $, $ \boldsymbol{A}_{ij}\in  \mathbb{R}^{n_i \times n_j}$, $ \boldsymbol{A}_D $ is a block-diagonal matrix, whose diagonal contains $ (\boldsymbol{A}_1, \cdots, \boldsymbol{A}_N) $, and $ \boldsymbol{A}_C $ is designed such that $ \boldsymbol{A}=\boldsymbol{A}_D+\boldsymbol{A}_C $. Given a potential attack on each subsystem, we can write the dynamic representation of a subsystem $i$: 
\begin{align}
\boldsymbol{E}_i\dot{\boldsymbol{x}}_i&= \boldsymbol{A}_i\boldsymbol{x}_i+\sum_{j\in \mathcal{N}^{\textrm{in}}_i}\boldsymbol{A}_{ij}\boldsymbol{x}_j,\\
\boldsymbol{y}_i&=\boldsymbol{C}_i\boldsymbol{x}_i, \textrm{ for } \hspace{0.3 cm} i\in \{1,\dots,N\},
\end{align}
where $ \boldsymbol{x}_i $ and $ \boldsymbol{y}_i $ are the state variables and sensor measurements of subsystem $ i $ and $ \mathcal{N}^{\textrm{in}}_i $ is the set of subsystems that depend on subsystem $ i $. As such, the distributed attack detection filter can be expressed as  \cite{pasqualetti2013attack}:
\begin{equation}
	\begin{aligned} \label{distributed filter}
		\boldsymbol{E}\dot{\boldsymbol{z}}(t)&=\left(\boldsymbol{A}_D+\boldsymbol{G}\boldsymbol{C}\right)\boldsymbol{z}(t)+\boldsymbol{A}_C\boldsymbol{z}(t)-\boldsymbol{G}\boldsymbol{y}(t),
		\\
		\boldsymbol{r}(t)&=\boldsymbol{y}(t)-\boldsymbol{C}\boldsymbol{z}(t).
	\end{aligned} 
\end{equation}

In \eqref{distributed filter}, since matrices $ \boldsymbol{A}_D $, $ \boldsymbol{C} $, and $ \boldsymbol{G} $ are block diagonal, each subsystem must receive $ \boldsymbol{A}_C\boldsymbol{w}(t) $ from the connected subsystems to be able to compute the residue of the filter, continuously. To reduce the load and overhead that can result from a continuous communication between subsystems, we use the waveform relaxation approach in \cite{pasqualetti2013attack}. In this approach, to observe the existence of an attack during a time interval $ [0,T] $ where $ T>0 $, each subsystem $ i $ transmit its own states $ \boldsymbol{z}_i(t) $ to the subsystems in $ \mathcal{N}^{\textrm{out}}_i $ (the set of subsystems on which subsystem $ i $ is dependent) and will receive states $ \boldsymbol{z}_j $ from subsystems in $ \mathcal{N}^{\textrm{in}}_i$. Then, the waveform relaxation iteration for distributed filter can be written as follows:
\begin{equation}
	\begin{aligned}\label{localfilter}
		\boldsymbol{E}_i\dot{\boldsymbol{z}}_i^{(k)}(t)=
		\\(\boldsymbol{A}_i+\boldsymbol{G}_i\boldsymbol{C}_i)&\boldsymbol{z}_i^{(k)}(t)+\sum_{j\in\mathcal{N}_i^\textrm{in}}\boldsymbol{A}_{ij}\boldsymbol{z}_j^{(k-1)}(t)-\boldsymbol{G}_i\boldsymbol{y}_i.
	\end{aligned}
\end{equation}

Each subsystem has an initial estimation $ z_j^{(0)}(t) $ for the connected subsystems and using the following steps, it can find the local residue of filter \eqref{distributed filter} starting from step $ k=0 $:
\begin{enumerate}
	\item Increment $ k $ by one and compute $ z_i^{(k)}(t) $ from filter \eqref{localfilter}.
	\item Transmit $ z_i^{(k)}(t) $ to the subsystem $ j\in\mathcal{N}_i^\textrm{out}, $
	\item Receive $ z_j^{(k)}(t) $ from subsystems $ j\in\mathcal{N}_i^\textrm{in}$ and update $ w_j^{(k)} $,
\end{enumerate}
The residue in local filters for sufficiently large $ k=\bar{k} $ converge and can be used for local attack detection \cite{pasqualetti2013attack}.

The main shortcoming of this method is that each subsystem should wait for $ T $ seconds before starting to calculate the residue of its filter. This will make the system vulnerable to attacks for $ T $ seconds. To overcome this challenge, we next propose a game-theoretic approach.
\section{Game-Theoretic Attack Detection} \label{gamemod}
\subsection{Communication Limitation in Attack Detection} 
The presented distributed detection filter in Section \ref{defmod} does not account for the potential limitation on the communication capacity. This limitation bounds the needed communication of data between the subsystems for the distributed computation of the residues. In addition, the presented detection filter does not prioritize between the different subsystems based on their induced effect on the operation of the whole system or their level of vulnerability to attacks. The designer of the attack detection filters who is the administrator of the system, can reduce the time needed to detect an attack on a certain state by having more frequent communications between the subsystems. However, the limitations in communication resources such as bandwidth or processing power and the associated overhead prevents the defender from infinitely increasing the rate of communication among the subsystems to decrease the detection time to zero. In this regard, we consider that the total number of successful connections that the communication layer can provide is $ M $ per second. As such, we assume that the number of connections during $ T $ seconds for a subsystem $ i $ is $ m_i $. Here, the total number of connections cannot exceed $ M $ as captured by:
\begin{align}\label{constraint}\vspace{-0.3cm}
\sum_{i=1}^N m_i \leq T M.
\end{align}

In this respect, increasing the number of connections allocated to a certain subsystem can reduce the time over which this subsystem can be subject to an attack before attack detection. To this end, the defender can use a strategy which consists of assigning a time division $ m_i $ to each of the $ N $ subsystems  to decrease the potential damage inflicted by the attacker. Indeed, allocating $m_i$ connections to a subsystem $i$ allows the attacker to stay undetected for $ T/m_i $ seconds while attacking the states of $ i $, therefore, the defender can choose a higher value for $ m_i $ for the subsystems with higher vulnerabilities to the state attacks to reduce the detection delay on those subsystems. Considering $ \mathcal{M} $ as the set of possible communication resource allocations between the different subsystems, we can rewrite \eqref{pricechange} as follows:\vspace{-0.3cm}
\begin{equation}
	\begin{aligned}\label{attackdamage}
		\Delta p(\kappa,\mu)=&\sum_{j \in \kappa}\int_{0}^{\frac{T}{m_j}}\mathscr{L}^{-1}\Bigg\{\Bigg(\sum_{i=1}^{n}(-1)^{i+j}c_{p_i}\times \\
		&\frac{|(s\boldsymbol{E}-\boldsymbol{A})_{\bar{i},\bar{j}}|}{|s\boldsymbol{E}-\boldsymbol{A}|}\Bigg)v_j(s) \Bigg\}dt,
	\end{aligned}
\end{equation}
where $ \mu =\{m_1,\dots,m_N\} \in \mathcal{M} $. \eqref{attackdamage} represents the deviations in the operational states of the system (i.e. electricity costs) achieved by the attacker by choosing the attack set $ \kappa $ while the set of allocated communication resources by the defender is $ \mu $. Since the attacker aims to increase the cost deviation by selecting proper states to attack and the defender seeks to reduce the deviation by optimal allocation of the communication resources on the subsystems, we can use a \emph{game-theoretic approach} to solve this problem \cite{bacsar1998dynamic}.\vspace{-0.1cm}
\subsection{Game Formulation and Solution}
To model the interdependent decision making process of the attacker and defender, we introduce a zero-sum noncooperative game in strategic form $ \left\{\mathcal{Q},\left\{\mathcal{S}_{i}\right\}_{i \in\mathcal{Q}},\left\{u_i\right\}_{i\in\mathcal{Q}}\right\} $ defined by three components: a) the \emph{players} which are the attacker $ a $ and defender $ d $ in the set $ \mathcal{Q}:=\left\{a,d\right\} $, b) the \emph{strategy} spaces $ \mathcal{S}_i $ for $i\in\mathcal{Q}$, and c) the \emph{utility function} $ u_i $ of each player.

For the attacker, the set of strategies $\mathcal{S}_a$ correspond to the set of states to attack simultaneously among the different states of the interconnected infrastructure. Hence, $\mathcal{S}_a$ is defined as:
\begin{align}\label{attackset}
 \mathcal{S}_a\triangleq\mathcal{K}=\left\{\kappa\big|\kappa=\left\{ b_j\in \left\{0,1\right\} \Bigg|\sum _{j=1}^nb_{j}\leq K \right\}\right\},
\end{align}
where $ K $ is the number of states  that the attacker can attack simultaneously and $ n $ is the number of states of the entire interdependent system. For the defender, given the limitation on the traffic load, it can assign different observation periods for each subsystem by assigning $ m_i $ to each subsystem while meeting the constraint in \eqref{constraint}. Therefore, the strategy set of the defender consists of the different possible allocations of the communication resources among the subsystems:
 \begin{align}\label{defenseset}
  \mathcal{S}_d\hspace{-0.1cm}\triangleq\hspace{-0.1cm}\mathcal{M}=\hspace{-0.1cm}\left\{\mu\big|\mu=\hspace{-0.1cm}\left\{\hspace{-0.1cm}m_i\in\hspace{-0.1cm} \left\{1,\dots,TM\right\} \hspace{-0.1cm}\Bigg|\hspace{-0.1cm}\sum _{i=1}^{N}m_i\hspace{-0.1cm}\leq \hspace{-0.1cm}TM\hspace{-0.1cm} \right\}\hspace{-0.1cm}\right\}\hspace{-0.1cm}.
 \end{align}
 For a chosen attack strategy $\kappa \in\mathcal{S}_a$ and $\mu\in\mathcal{S}_d$, the utility function of the attacker and defender corresponds to: 
\begin{align}
u_a(\kappa,\mu)=-u_d(\kappa,\mu)= \Delta p(\kappa,\mu). 
\end{align}
\subsection{Mixed Strategy Nash Equilibrium Solution}
For the studied attack detection game, players can choose one of the strategies in their strategy set or assign probabilities for playing each of the strategies which is called the \emph{mixed strategy}\cite{bacsar1998dynamic}. Mixed strategies are motivated by two facts: a) both players must randomize over
their strategies in order to make it nontrivial for the opponent to guess their potential actions, and b) the communication resource allocation and state attacks can be repeated over an infinite time duration, and therefore mixed strategies can capture the frequency of choosing certain strategies for both players. Let $ \boldsymbol{p}_a $ be the vector of \emph{mixed strategies} for the attacker where each element in $ \boldsymbol{p}_a $ is the probability of selecting a set of states $ \kappa \in \mathcal{S}_a $ to attack simultaneously and $ \boldsymbol{p}_d $ be the vector of mixed strategies for the defender whose elements represent the probability of allocating a certain amount of resources on the subsystems as captured by $ \mu \in \mathcal{S}_d $.

In a game-theoretic setting \cite{bacsar1998dynamic}, each player chooses its own mixed-strategy vector to maximize its expected utility. The utility of each player is the expected value over its mixed strategies, which for any of the two players $ i\in \mathcal{Q} $, is: 
\begin{align}\label{expectedutility}
	U_i(\boldsymbol{p}_a,\boldsymbol{p}_d)=\sum_{\boldsymbol{s} \in \mathcal{S}}\left(p_d\left(\mu\right)p_a\left(\kappa\right)\right)u_i(\boldsymbol{s}),
\end{align}
where $ \boldsymbol{s}=[\mu \; \kappa] $ is a vector of selected pure strategies and $ \mathcal{S}=\mathcal{S}_a \times \mathcal{S}_d $. In our game, the defender seeks to minimize \eqref{expectedutility} while the attacker seeks to maximize it.

To solve the proposed game, we seek to find the mixed-Nash equilibrium, defined as follows:
\begin{definition}
A mixed strategy profile $ \boldsymbol{p}^* $ constitutes a \emph{mixed strategy Nash equilibrium} if for the defender, $ d $, and attacker, $ a $, we have:
\begin{equation}
	\begin{aligned}
		U_d(\boldsymbol{p}^*_d,\boldsymbol{p}^*_a)\leq U_d(\boldsymbol{p}_d,\boldsymbol{p}^*_a),\quad \forall \boldsymbol{p}_d \in \mathcal{P}_d,\\
		U_a(\boldsymbol{p}^*_d,\boldsymbol{p}^*_a)\geq U_a(\boldsymbol{p}^*_d,\boldsymbol{p}_a),\quad \forall \boldsymbol{p}_a \in \mathcal{P}_a,
	\end{aligned}
\end{equation}
where $ \mathcal{P}_i $ is the set of all probability distributions for player $ i $ over its action space $ \mathcal{S}_i $.
\end{definition}
It is well-known that there exists at least one mixed-strategy Nash equilibrium (MSNE) for any finite noncooperative game\cite{bacsar1998dynamic}. The MSNE for our game implies a state at which the defender has chosen its optimal randomization over the resource allocations and, therefore, cannot improve its utility by changing this allocation. Similarly, for the attacker, an MSNE for attacker represents a state at which the attacker has chosen its optimal randomization over the selection of states to initiate an attack and, thus, cannot improve its utility by changing this choice. Since our game is a zero-sum two-player, to find a closed-form solution for the MSNE, we can use the von Neumann indifference principle \cite{bacsar1998dynamic}. Under this principle, the expected utilities of players at MSNE for of any pure choice under the mixed strategies played by the opponent, must be equal. Given the large strategy space for the players in our game, solving the equations derived from von Neuman approach can be challenging, thereby, we use a learning algorithm named \emph{fictitious play} \cite{rose2011learning}, which is known to converge to an MSNE for any two-player zero-sum game.
\begin{figure}[!t]
	\begin{center}
		\vspace{-0.1cm}
		\includegraphics[width=.75\columnwidth]{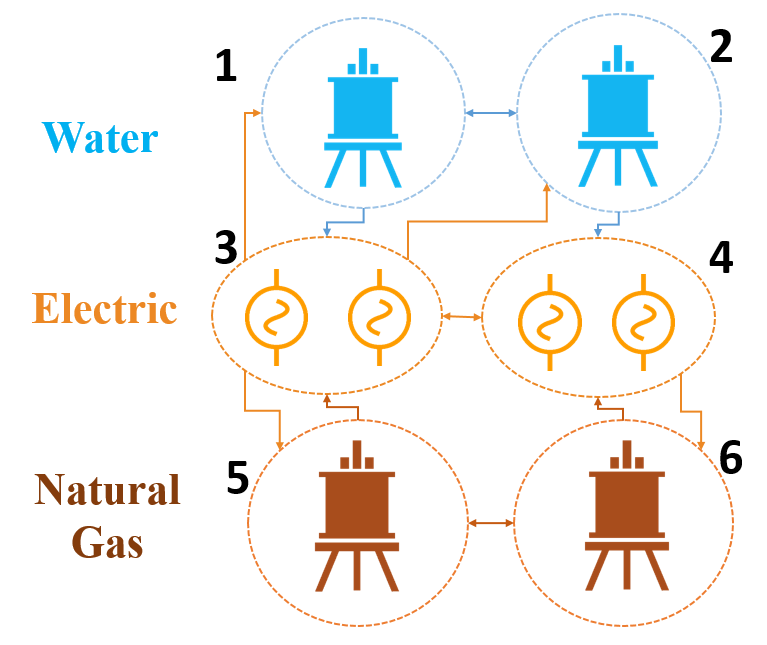}
		\vspace{-0.2cm}
		\caption{Example of interconnected infrastructure: the blue, orange, and brown elements are the components of the water, electric and natural gas system.}
		\label{system}
	\end{center}\vspace{-0.6cm}
\end{figure} 
 \begin{figure}[!t]
	\begin{center}
		\vspace{-0.1cm}
		\includegraphics[width=0.9\columnwidth]{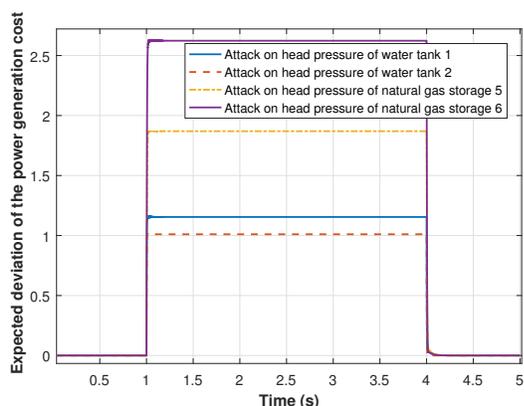}
		\vspace{-0.2cm}
		\caption{Percentage of deviation from power generation cost resulting from the attack on the states of the gas and water system. Attacks initiated at $ t=1 $s and lasted until $ t=4 $s.}
		\label{pricedeviation}
	\end{center}\vspace{-0.75cm}
\end{figure} 
\section{Simulation Results}\label{sim results}
In this section, we simulate a system of interconnected critical infrastructure having two electric subsystems, two water subsystems, and two natural gas subsystems consisting of four generators, two tanks, and two gas storage units, respectively, as shown in Fig. \ref{system}. In this model, we have 8 electric system states (4 generator phase angle and 4 angular speed), 2 water system states and 2 natural gas states. Therefore, 12 states in the interconnected system can be attacked by the attacker. The physical parameters of the system are chosen such that \eqref{GeneralDAE} is asymptotically stable. We observe the system for 5 seconds during which the defender can allocate communication resources over 6 subsystems. For the proposed game, we assume that the highest number of the states that the attacker can choose is 5 and we simulate the system for different numbers of available communication resources for the defender. To obtain the MSNE under different conditions, we use the algorithm proposed in \cite{rose2011learning}.

We simulated the attacks on the states of the natural gas and water systems during the time interval of $ [1,4] $ seconds without any resource allocation on the subsystems to show the deviation in power cost caused by attacking to the natural gas and water systems. Fig. \ref{pricedeviation} shows that the dependence of the power system on water and natural gas provides an opportunity for the attacker to change the power cost by attacking the water and natural gas systems. From Fig. \ref{pricedeviation} we can see that, when the attacker initiates an attack on one of the states of the water or natural gas system at $ t=1 $s the power generation cost starts to deviate and the deviation lasts until the attacker stops the state attack at $ t=4 $s.

\begin{figure}[!t]
	\begin{center}
		\vspace{-0.1cm}
		\includegraphics[width=0.85\columnwidth]{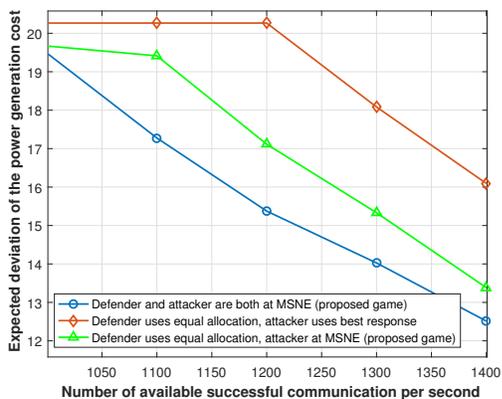}
		\vspace{-0.2cm}
		\caption{Expected cost deviation as function of communication resources.}
		\label{EqualAllocation}
	\end{center}\vspace{-0.6cm}
\end{figure}

In Fig. \ref{EqualAllocation}, we show the expected cost deviation as a function of available communication resources. Fig. \ref{EqualAllocation} illustrates that, the expected utility for the attacker decreases by 36\% with 40\% increase in the communication resources when both the defender and attacker play their MSNE. To show the benefits of using a game-theoretic solution, we compare with the case in which the defender equally allocates the communication resources on the subsystems without strategic behavior, while, attacker: 1) plays the MSNE obtained from the previous simulation or 2) plays the pure strategy which maximizes its expected utility, i.e., the \emph{best response}. Fig. \ref{EqualAllocation} shows that the expected deviation of the cost when the defender allocates resources equally over the subsystems, is 35\% higher than when the defender plays the MSNE. This illustrates that using the proposed game-theoretic solution, the defender can reduce the expected cost deviation. In addition, Fig. \ref{EqualAllocation} shows that, while the defender equally allocates the resources, attacker can increase the cost deviation by 18\% if it chooses to play the best response instead of MSNE. Under equal allocation, since the defender allocates the resources without considering the attacker's strategy, the attacker does not randomize between the strategies and chooses the strategy which maximizes its utility and, therefore, the best response of the attacker to equal allocation of resources yields a higher utility for the attacker.

\begin{figure}[!t]
	\begin{center}
		\vspace{-0.1cm}
		\includegraphics[width=0.85\columnwidth]{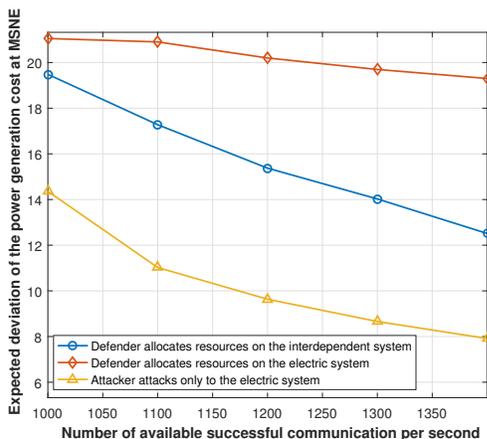}
		\vspace{-0.3cm}
		\caption{Comparison of the expected cost considering interdependent and disjoint systems.}
		\label{MSNEcomparison}
	\end{center}\vspace{-0.6cm}
	\end{figure}

In Fig. \ref{MSNEcomparison}, we compare two cases: a) a first case in which the defender only protects the electric system, b) a second case in which the defender protects all the interdependent systems. From Fig. \ref{MSNEcomparison}, we can see that, if the defender allocates the resources only on the electric system, the expected cost deviation increases by 30\%, approximately. From the attacker's perspective, the attacker can increase the expected cost deviation by 35\% if it is able to attack the states of the natural gas and water systems compared to the case in which the attacker attacks only the states of the power system. Therefore, we can conclude that the interdependence between the electric, natural gas, and water systems can potentially increase the system's vulnerability to attacks.

Finally, to provide further insight on the allocation of communication resources and selection of states to attack, we consider a scenario in which the available communication resources for the defender are equal to $ M=1200 $ and the attacker can attack up to $ K=5 $ states. The MSNE for the attacker in this case is mixing between $ \{\delta_{1_1},\omega_{2_2},h_{s_1}^G,h_{s_2}^G,h_{s_2}^W,\} $ and $ \{\delta_{1_1},\delta_{1_2},\omega_{2_1},h_{s_1}^G,h_{s_2}^G,\} $ with the probability vector $ \boldsymbol{p}_a=[0.872,0.128] $ while the defender allocates the resources to 6 subsystems as $ \{m_1=300,m_2=400,m_3=100,m_4=200,m_5=100,m_6=100\} $ and $ \{m_1=300,m_2=300,m_3=200,m_4=200,m_5=100,m_6=100\} $ with the probability vector $ \boldsymbol{p}_d=[0.0183,0.9817] $ in its MSNE. From this result, it is clear that, at the MSNE, the attacker tends to attack to some of the water and natural gas states and, also, the defender allocates some of its communication resources to the water and gas subsystems to prevent the deviations in the power generation cost.
\section{Conclusion}\label{conclusion}
In this paper, we have proposed a novel game-theoretic approach for modeling the interactions between the administrator of an infrastructure and an adversary. The administrator seeks to observe the operation of the system by allocating communication resources over the local subsystems of its infrastructure. Meanwhile, the adversary seeks to attack the states of the infrastructure to disrupt the system's nominal operation. ٌWe have formulated the problem as a noncooperative zero-sum game where the defender allocates limited communication resources among the subsystems of the interdependent system while the attacker tries to optimally select the states to attack. Our results have shown that the interdependence between electric, water, and gas systems makes the electric system vulnerable to state attacks in the natural gas and water system. To overcome this vulnerability, the defender needs to allocate some of the available communication resources to the natural gas and water infrastructure.
\bibliographystyle{IEEEtran}
\bibliography{references}
	
	
	

\end{document}